\documentclass[prb,superscriptaddress,showpacs,twocolumn,aps,a4paper,amsmath,amssymb]{revtex4}
\usepackage{dcolumn}
\usepackage{amsmath}
\usepackage{graphicx}
\usepackage{latexsym}
\usepackage{amsfonts}
\usepackage{amssymb}
\usepackage{bm}

\DeclareGraphicsExtensions{.pdf,.gif,.jpg}

\def\a{\alpha}
\def\b{\beta}
\def\e{\epsilon}
\def\ve{\varepsilon}
\def\d{\delta}

\def\g{\gamma}
\def\G{\Gamma}
\def\l{\lambda}

\def\w{\omega}

\def\bra{\langle}
\def\ket{\rangle}

\newcommand{\be}{\begin{equation}}
\newcommand{\ee}{\end{equation}}
\newcommand{\beq}{\begin{eqnarray}}
\newcommand{\eeq}{\end{eqnarray}}

\begin{document}


\title{Transient dynamics in the Anderson-Holstein model with
interfacial screening}

\author{E. Perfetto}

\affiliation{Dipartimento di Fisica, Universit\`a di Roma
 Tor Vergata, Via della Ricerca Scientifica 1, I-00133 Rome, Italy}
 \affiliation{INFN, Laboratori Nazionali di Frascati, Via E. Fermi 40,
 00044 Frascati, Italy}
 
\author{G. Stefanucci}

\affiliation{Dipartimento di Fisica, Universit\`a di Roma
 Tor Vergata, Via della Ricerca Scientifica 1, I-00133 Rome, Italy}
 \affiliation{INFN, Laboratori Nazionali di Frascati, Via E. Fermi 40,
 00044 Frascati, Italy}
\affiliation{European Theoretical Spectroscopy Facility (ETSF)}

\begin{abstract}
    
We study the combined effects of electron-phonon coupling
and  dot-lead repulsion in the transport 
properties of the Anderson-Holstein model.
We employ a recently proposed nonperturbative method to
calculate the transient response of the system.
By varying the initial conditions for the time propagation,
we are able to disentangle two different dynamical
processes, namely the local charge rearrangement due to the dot-lead contacting 
and the establishment of the nonequilbrium many-body state due to  
the application of the external bias. According to the 
distinct initial contacting, the current can exhibit transient oscillations
of different nature. These origin from tunneling events that
involve virtual Franck-Condon excitations, or virtual transitions between
the resonant level and the Fermi energy of the leads.  

\end{abstract}

\maketitle

\section{Introduction}

The interplay of vibrational and electronic degrees of freedom
during the charge tunneling through molecular junctions gives rise to
striking nonequilibrium effects\cite{troisi}. 
These include
phonon-assisted tunneling\cite{phonass1,phonass2},
hysteresis-induced bistability\cite{bist1,bist2,bist3}, local 
heating\cite{heat}, molecular 
switching\cite{switch1,switch2}, and negative differential 
conductance\cite{phonass2,ndrexp1,ndrexp2}.
Furthermore, if the electronic coupling to the vibrations is particularly strong
a collective phenomenon known as Franck-Condon blockade (FCB) 
occurs\cite{fcb}. This manifests itself in a dramatic suppression of the tunneling
current at low bias, accompanied  by the appearance of several
vibrational sidebands inside the Coulomb-blockade diamonds,
recently observed in carbon nanotubes quantum dots (QD)\cite{vonoppen}.

From the theoretical side, the quantitative description and 
understanding of the above phenomena represents a challenging nonequilibrium problem.
Many observed properties can be addressed
within the Anderson-Holstein model\cite{wingreen}, that describes a single electronic level
coupled linearly to a vibrational mode and to  metallic electrodes.
Despite its apparent simplicity, however, this model
does not allow for an analytic solution
and approximate approaches must be resorted.
Only very recently, numerically exact methods
to calculate its nonequilibrium properties have been
put forward\cite{rabani,albrecht,wilner,thossexact,wilner2}. Beside
providing a valuable validation of previous findings, 
the exact data have also revealed novel features, especially in the 
time domain, like the extraordinarily long-transient
dynamics needed to reach the stationary state in the FCB regime\cite{albrecht}.

The inclusion of electron correlations widens the range of accessible phenomena,
but at the same time 
complicates the theoretical treatment even further. 
While a considerable amount of work has been devoted to
study intra-molecule electron-electron ({\it e-e}) 
interactions\cite{review} (mainly to  address the
joint effects of vibrations and Coulomb blockade or
Kondo-like correlations), the role of interfacial 
repulsion between the molecule and the leads
has been explored only very recently\cite{perfetto1}.
In that paper we    
developed a novel approach to study 
the complex interplay between the local electron-phonon ({\it e-p}) coupling
together with the  
molecule-lead {\it e-e} repulsion.
We showed that the exciton formation at the molecule-lead interface
improves significantly the polaron mobility thus competing with the FCB.
As a consequence the FCB regime is dynamically established after a 
long-lasting sequence of blocking-deblocking
events\cite{albrecht}, characterized by exciton-enhanced current spikes.

In this paper we extend our previous study by computing the 
time-evolution of QD 
density under the application of an external bias, and by investigating
how the transient current is modified by changing the initial conditions
for the time propagation.
When possible, we also compare our results with the exact data, finding 
very good agreement.
We show that different ways of connecting in time the QD and the leads
can produce qualitative differences in the transient current, 
accompanied by a change of the nature of the dominant oscillations.

\section{Model and formalism}

The model we consider consists in a single-level QD attached to two 
semi-infinite one-dimensional noninteracting wires. An electron occupying the level 
is coupled to a single vibrational mode located in the QD, and 
at the same time interacts with the electrons in the leads which are in the proximity 
of the QD. The spinless Hamiltonian describing this  
system is given by (in standard notation)
%
\beq
\hat{H}&=&-t_{w}\sum_{\a,j=1}^{\infty}(\hat{d}^{\dagger}_{\a j}\hat{d}_{\a 
j+1}+\mathrm{h.c.})
+ T_{l} \sum_{\a}( \hat{d}^{\dag}_{\a 1}\hat{d} +\mathrm{h.c.} ) 
\nonumber \\
&+& \e_{d} \hat{n}_{d}+ \omega_{0}\hat{a}^{\dagger}\hat{a}  + 
\lambda \hat{n}_{d} (\hat{a}^{\dag}+\hat{a})  +
U \hat{n}_{d}\sum_{\a}\hat{n}_{\a 1}, \; \; \; \;
\label{eq1}
\eeq
where $\a=L,R$ labels the Left and Right lead, and
$\hat{n}_{d}=\hat{d}^{\dag}\hat{d}$, $\hat{n}_{\a 0}=
\hat{d}^{\dagger}_{\a 0}\hat{d}_{\a 0}$ are the 
densities on the 
QD and on the first sites of the lead $\a$ respectively.
The system is perturbed by  an 
external bias given by
$\hat{H}_{V}= \sum_{\a} V_{\a} \hat{N}_{\a}$, with
$\hat{N}_{\a}=\sum_{x} \hat{n}_{\a x} $ the total number of 
particles in wire $\a$ and $V=V_{L}-V_{R}$ the total 
applied voltage.

In the following we consider the continuum version $\hat{H}_{c}$ of the  
above model  since it allows for a nonperturbative treatment of 
the {\it e-e}  interaction via the bosonization 
technique\cite{giamarchi}.
To this end we first assume half-filled wide 
band leads with linear dispersion $\e_{k}=v_{F}k$ (with $v_{F}=
2t_{w} a$  
the Fermi velocity and $a$ the lattice spacing) and constant tunneling amplitude 
$\G=2\pi T_{l}^{2} \sum_{k}\delta(\omega-\e_{k})=2T_{l}^{2}/t_{w}$, and then we unfold the 
left and right leads\cite{boulat,perfetto1}.
In this way the first term of Eq. (\ref{eq1}) takes the Dirac-like 
form $-\sum_{\a} i v_{F} \int dx \, \hat{\psi}^{\dagger}_{\a}(x) 
\partial_{x}\hat{\psi}_{\a}(x)$,
where the electron field operator $\hat{\psi}_{\a}(x)$ describes
an electron at position $x$ in the (chiral) lead $\a$ moving with
velocity $v_{F}$. The rest of the continuum model is simply obtained by
replacing $\hat{d}_{\a x} \to  \hat{\psi}_{\a}(x)$, $\sum_{x} \to \int dx$,
and by rescaling the model parameters according to $T_{l}\to t_{l}=2\sqrt{a}T_{l}$ and
$U\to u=4aU$.
We then bosonize the electron operators as \cite{giamarchi}
$
\hat{\psi}_{\a}(x)=\frac{\eta_{\a}}{\sqrt{2\pi a}}
e^{-2\sqrt{\pi}\,i\hat{\phi}_{\a}(x)},
\label{bospsi}
$
with boson field $
\hat{\phi}_{\a}(x)= i \a \sum_{q>0}\zeta_{q}
(\hat{b}^{\dagger}_{\a q }e^{-i\a qx}-\mathrm{h.c.})
- \sqrt{\pi}x\hat{N}_{\a}/ \mathcal{L}
$, and $\eta_{\a}$ the anticommuting Klein factor. In the mode expansion of 
the boson field it holds $\zeta_{q}=\frac{ e^{- \frac{av_{F}q}{2}} 
}{\sqrt{2\mathcal{L}q}}$, with $\mathcal{L}$ the length of the system.
The electron density in the leads takes the form 
$
\hat{n}_{\a}(x)=-\partial_{x}\hat{\phi}_{\a}(x)/\sqrt{\pi}$,
and hence, up to an irrelevant term\cite{irlm}, 
the bosonized continuum Hamiltonian reads\cite{nota}
\beq
\hat{H}_{c}&=&\sum_{\a , q>0} v_{F} q \hat{b}^{\dagger}_{\a q} 
\hat{b}_{\a q} +\varepsilon_{d}\hat{n}_{d} 
+ \omega_{0}\hat{a}^{\dagger}\hat{a}
\nonumber\\
&+&
t_{l} \sum_{\a}
\left[
\frac{\eta^{\dagger}_{\a}}{\sqrt{2\pi} }
e^{-2\sqrt{\pi}\sum_{q>0}\zeta_{q}
(\hat{b}^{\dagger}_{\a q}- \hat{b}_{\a q})} \hat{d} +  \mathrm{h.c.}
\right] \nonumber \\
&+&  \hat{n}_{d} \left[ 
\lambda  (\hat{a}^{\dag}+\hat{a}) -u\sum_{\a ,q>0}
\frac{\zeta_{q}q}{\sqrt{\pi}}\,
(\hat{b}^{\dagger}_{\a q} +\hat{b}_{\a q}) \right].
\label{hboson}
\eeq
Next we perform a multi-boson Lang-Firsov transformation to (formally) eliminate 
the {\it e-p} and {\it e-p} coupling appearing in the last 
line of the above equation\cite{irlm,perfetto1}.
The unitary operator
$
\hat{\mathcal{U}}=\mathrm{exp}[-\frac{\l}{\w_{0}}
(\hat{a}^{\dag}-\hat{a})+2\sqrt{\pi}u\sum_{\a q} 
\frac{\zeta_{q}}{2\pi v}
(\hat{b}^{\dagger}_{\a q }-\hat{b}_{\a q}  ) ]\hat{n}_{d}
$
transforms the continuum Hamiltonian into 
$\hat{H}_{c}'=\hat{\mathcal{U}}^{\dagger}\hat{H}_{c} 
\hat{\mathcal{U}}$ with (from now on all sums are over $q>0$)
\be
\hat{H}_{c}'=\sum_{\a q} v_{F} q \hat{b}^{\dagger}_{\a q}
\hat{b}_{\a q} + \omega_{0}\hat{a}^{\dagger}\hat{a}
+\tilde{\varepsilon}_{d}\hat{n}_{d}  
+
t_{l}\sum_{\a}
\left[ \hat{f}^{\dag}_{\a 0} \hat{d} +  \mathrm{h.c.} 
\right] .
\label{hambos}
\ee
In the transformed Hamiltonian it appears the renormalized fermion field  
\be
\hat{f}_{\a x}=\frac{\eta_{\a}}{\sqrt{2\pi a}}
e^{-\frac{\l}{\w_{0}}(\hat{a}^{\dag}-\hat{a})+ 2\sqrt{\pi}\sum_{\b 
q}\zeta_{q}  W_{\a \b }
(\hat{b}^{\dagger}_{\b q} e^{-i qx}- \hat{b}_{\b q} e^{i qx})}
\ee
evaluated in $x=0$, with 
the effective interactions
$W_{RR}=W_{LL}=1-u/(2\pi v_{F} )$ and 
$W_{RL}=W_{LR}=-u/(2\pi v_{F} )$, and  renormalized  energy  level
$\tilde{\ve}_{d}=\ve_{d}-\frac{\l^{2}}{\w_{0}}-u^{2}\sum_{q}
\frac{e^{-aq}}{\pi v_{F} \mathcal{L}}$.
In the new basis we have a noninteracting QD coupled to  
effective leads (bosonic baths) whose ground state for $t_{l}=0$ is $|\Psi_{0} 
\ket=| 0_{p} \ket \otimes \prod_{\a q}|0_{\a q}\ket  $, 
where $| 0_{p} \ket $ and $|0_{\a q}\ket$ are the vacua of the
boson operators  $\hat{a}$ and $\hat{b}_{\a q}$ respectively; the tunneling 
coupling occurs via the correlated-polaron operator $\hat{f}$.

\section{Equation of motion}

The great advantage of casting the Hamiltonian in the form of
Eq. (\ref{hambos}) is the possibility of 
writing an approximate equation of motion for the nonequilibrium QD Green's 
function, solely in presence of a {\it correlated-polaron} embedding 
self-energy. The latter accounts for the presence of the biased 
leads as well as for the {\it e-p} and {\it e-p} interactions 
in a nonperturbative way. We define the QD Green's function on 
the Keldysh contour as
$G(z,z')=\frac{1}{i} \bra {\cal T} {\hat{d}(z) \hat{d}^{\dag}(z')} \ket,$
where ${\cal T}$ is the contour ordering, operators are in the 
Heisenberg picture with respect to 
$\hat{H}_{c}'+\hat{H}_{B}$ (the bias perturbation does not change after the 
transformation); the average is taken over 
the uncontacted ground state $|\Psi_{0}\rangle\otimes |n_{d} \rangle$, 
$|n_{d}\rangle$ being the state of the QD 
with density $n_{d}$, with $n_{d}$ ranging from $0$ to $1$. In Ref. 
\onlinecite{perfetto1} we proposed a controlled 
approximation scheme in order to derive a closed (and numerically solvable)
equation of motion for $G$, that reads
\be
(i\partial_{z}-\tilde{\ve}_{d})G(z,z') - \int_{\g} d\bar{z} \sum_{\a} 
\Sigma_{\a}(z,\bar{z}) G(\bar{z},z')=\d(z,z') ,
\label{eom4}
\ee
where
$
\Sigma_{\a}(z,z') 
$
is the  correlated-polaron embedding self-energy. The 
real-time Keldysh components of $\Sigma$
can be evaluated exactly using again the bosonization method\cite{giamarchi}
and read
\beq
\Sigma^{\lessgtr}_{\a}(t,t')=\pm 
\frac{iv_{F}\G e^{-g}}{4\pi a^{1-\b}  } 
e^{i[\varphi_{\a}(t')-\varphi_{\a}(t)] }
\frac{ e^{ge^{\pm i\w_{0}(t-t')}}}{[a \mp iv_{F}(t-t')]^{\b}},
\label{sigmalutt}
\eeq
with adimensional {\it e-p} coupling $g=(\l/\w_{0})^{2}$, 
interaction dependent exponent  $\b=1+\frac{u(u-2\pi 
v_{F})}{2\pi^{2}v_{F}^{2}}$ and phase $\varphi_{\a}(t)=\int^{t}_{0} 
d\bar{t}\, V_{\a}(\bar{t})$.
The power-law reflects the collective excitonic response
of the lead electrons to the attractive potential due to the creation 
of a hole in the QD\cite{mahan}. 
In the noncorrelated case we have $\b=1$, while electron correlations 
produce $\beta<1$; the smaller the exponent is, the stronger is the exciton effect.

The integral in Eq. (\ref{eom4}) runs 
over the Keldysh contour $\g$, and
using the Langreth rules\cite{keldysh} 
it is converted into a coupled system of Kadanoff-Baym
equations\cite{dvl.2007,mssvl.2009} (KBE) which we solve 
numerically.
Once the Keldysh components of $G(z,z')$ are known,
the time-dependent QD density is calculated as
$n(t)=-iG^{<}(t,t)$, while the transient current
flowing through the QD and the $\a$ lead can be evaluated according 
to
\be
I_{\a}(z)=\int_{\g}d\bar{z} \,\Sigma_{\a}(z,\bar{z}) G(\bar{z},z) 
+\mathrm{h.c.}\quad
\label{tdcurr}
\ee

\section{Initial conditions}

\begin{figure}[tbp]
\includegraphics[width=7.25cm]{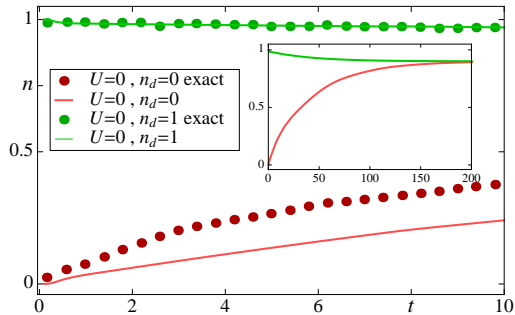}
\caption{TD density in the partitioned approach for $U=0$ with initial QD occupancy $n_{d}=0$ (red) and $n_{d}=1$ 
(green).
Exact data from Ref. 
\onlinecite{albrecht} are also displayed (circles). The rest of parameters are $\l=16$, $\w_{0}=8$, $V=26$, 
$\tilde{\e}_{d}=-10$, $v_{F}/a=100$. Units: $\G$ for energies 
and $\G^{-1}$ for times. The insets displays the TD 
result for a longer propagation time, in order to appreciate that the two densities reach 
the same steady-state value.  }
\label{fig1}
\end{figure}

Before presenting the explicit numerical results, a 
brief discussion about the initial conditions is in order.
To solve practically Eq. (\ref{eom4}), one has to set the initial value
$G^{<}(0,0)=-in_{d}$, that corresponds to (apart from the $-i$ factor) the initial 
density in the QD. This means that our propagation scheme assumes
the system initially {\it uncontacted} ($t_{l}=0$, i.e. the leads and 
the QD are in their ground states),
and the contacts and bias are switched at time $t=0$.
Therefore the subsequent transient regime accounts for two different
dynamical processes: (i) the charge rearrangement due to the QD-lead contacting 
(creation of Friedel-like oscillations in the leads, etc),
and (ii) the establishment of a genuine nonequilbrium many-body state due to  
external bias (the rise of the current towards its steady-state value, etc).
This scheme corresponds to the so-called {\it partitioned} approach.
However, in a more realistic situation, the bias is switched when the 
system is already contacted and in equilibrium (i.e. the leads and the
QD are in the ground state with $t_{l} \neq 0$).
This is the so-called {\it partition-free} approach\cite{cini,stefalmb}.
Here the transient dynamics can be very different from the 
partitioned case, since the two processes (i) and (ii) 
described above are not 
superimposed\cite{perfspin,perfspin2,perfgraph,perfspin,perfjosep,riku}.
In our scheme we can numerically simulate the partition-free appraoch,
since the bias function $V_{\a}(t)$ (appearing in the phase 
$\varphi_{\a}$ 
in Eq. (\ref{sigmalutt}))
is completely arbitrary.
In practice we consider a step-like bias function 
$V_{\a}(t)=V_{\a}\theta(t_{th})$, that corresponds to take the system initially
uncontacted with a given $n_{d}$ at time $t = 0$, let the system
thermalize (dynamics (i)) till a time $t_{th}$ at which no current flows across 
the links, and then we switch the bias perturbation on (dynamics 
(ii))\cite{gkba}.

\section{Transient density}

In Fig. \ref{fig1} we assess the accuracy of the proposed approach 
by comparing our results against exact data availabale in the 
literature for $U=0$, and obtained within the partitioned scheme\cite{albrecht}. 
It appears that the agreement is exceptionally good for initial 
density $n_{d}=1$, while for $n_{d}=0$ we predict a slower raise of the 
density towards its steady-state value. We recall, however, that in 
this case our results improve the state-of-the-art\cite{albrecht}.
In the inset  we show the TD 
density for a longer propagation time (not within reach of current numerical techniques),
in order to appreciate that the two densities reach the same 
steady-state value, as it should be.

\begin{figure}[tbp]
\includegraphics[width=7.25cm]{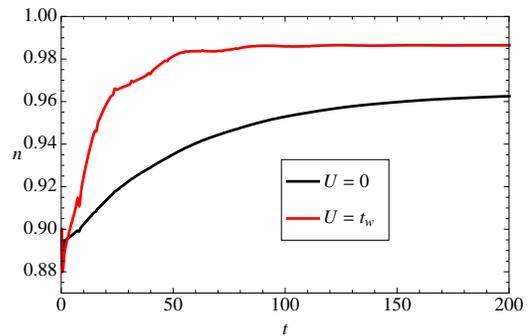}
\caption{TD relaxation towards the equilibrium density for $U=0$ (black) and $U=t_{w}$ (red), with initial QD 
occupancy $n_{d}=0.9$ and bias $V=0$. The rest of parameters and 
units are the same as in Fig. \ref{fig1}.   }
\label{fig2}
\end{figure}

We now study the effects induced by the QD-lead
repulsion $U$. Fig. \ref{fig2} displays 
the relaxation of the QD density
from the uncontacted value ($n_{d}=0.9$)
to the equilibrium value, after that the system has been contacted 
without bias.
We see that the effect of the screening interaction is twofold: 
enhance the asymptotic value of $n(t)$, and speed up
the relaxation time. 
The first effect origins from the fact that 
 a finite electron density in the QD induces
a charge depletion in the portion of the leads which are in the proximity 
of the interface; part of the repelled charge, in turn, migrates towards the 
QD thus enhancing its population. 
The second, instead, is a jamming effect reminiscent of the one 
found in Ref. \onlinecite{irlm}, that is due to the 
dynamical screening of the QD charge and 
tends to stabilize faster the value of $n(t)$.

\section{Transient current}

\begin{figure}[tbp]
\includegraphics[width=7.25cm]{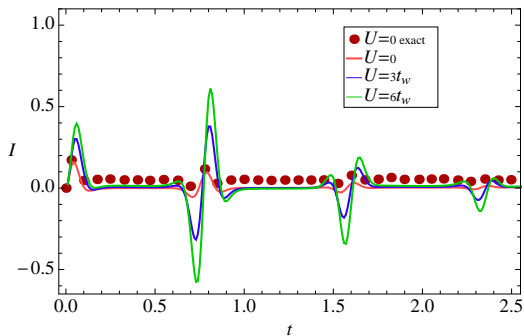}
\caption{TD current within the partitioned scheme for different $U$, 
initial QD occupancy $n_{d}=1$, and bias voltage $V=5\Gamma$.
For $U=0$, exact data from Ref. 
\cite{albrecht} are also displayed (circles). The rest of parameters and 
units are the same as in Fig. \ref{fig1}.}
\label{fig3}
\end{figure}

In this section we study the TD current $I(t)=[I_{L}(t)+I_{R}(t)]/2$
flowing under the application of the external bias.
As in the case of the density, the approach is first validated by 
comparing our results with the exact 
results recently obtained  for $U=0$ within diagrammatic Monte Carlo 
simulations\cite{albrecht}.
In Fig. \ref{fig3}  a remarkable agreement between the 
two approaches (within the partitioned scheme) can be appreciated.
In particular our method  
efficiently reproduces the very peculiar transient behavior of $I(t)$ 
before the steady-state is reached. 
In the absence of {\it e-e} interactions 
the time-dependent current displays quasi-stationary plateaus where 
almost no electron tunnels across the junction. At times 
$t_{n}=2n \pi / \w_{0}$ a deblocking effect occurs, and the current 
exhibits narrow bumps, signaling a sudden electron flow.
When the {\it e-e} interaction is considered, we observe a 
significant enhancement of the transport properties, characterized 
by larger current spikes. 
As we show below, the physical interpretation of 
such striking transient dynamics
provides a clue to understand how the FCB regime is dynamically 
established, and how {\it e-e} interaction modify the FCB scenario.
At $t=t_{n}$ an 
electron occupying the QD is in the polaron ground-state with  
phonon cloud centered at $x \sim n\l$ (with $n\approx 1$). At this time the polaron 
tunnels to the lead, causing a displacement of the oscillator to $x\to 0$ 
(since now $n=0$). At this point the polaron cannot hop back to 
the QD  since the overlap between the two shifted oscillator 
wavefunctions is negligible. Only after a vibrational period $2\pi / 
\w_{0}$ the overlap returns to be large, and the polaron can hop back to the 
QD.
Let us now consider the effects of {\it e-e} interaction.
The mechanism described above is modified as follows: If at time $t=t_{n}$
ane electron is on the QD, the electron density diminishes 
at the site of the lead boundary, thus 
overcoming the hopping suppression due to the Pauli principle 
and enhancing the effective tunneling rate\cite{borda,goldstein,irlm}.
Similarly, at time  $t=t_{n+1}$ the electron can easily tunnel back to the 
QD, being attracted by the hole previously left. 
This explains in a transparent way the $U$-induced enhancement of the current 
spikes observed in Fig \ref{fig3}.

\begin{figure}[tbp]
\includegraphics[width=7.25cm]{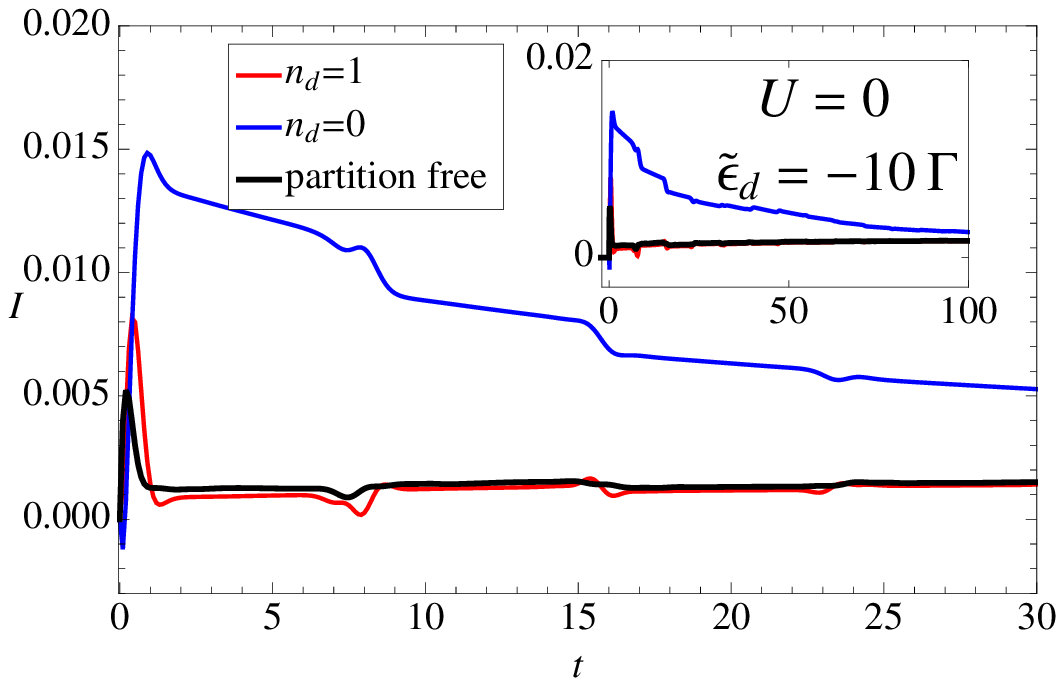}
\includegraphics[width=7.25cm]{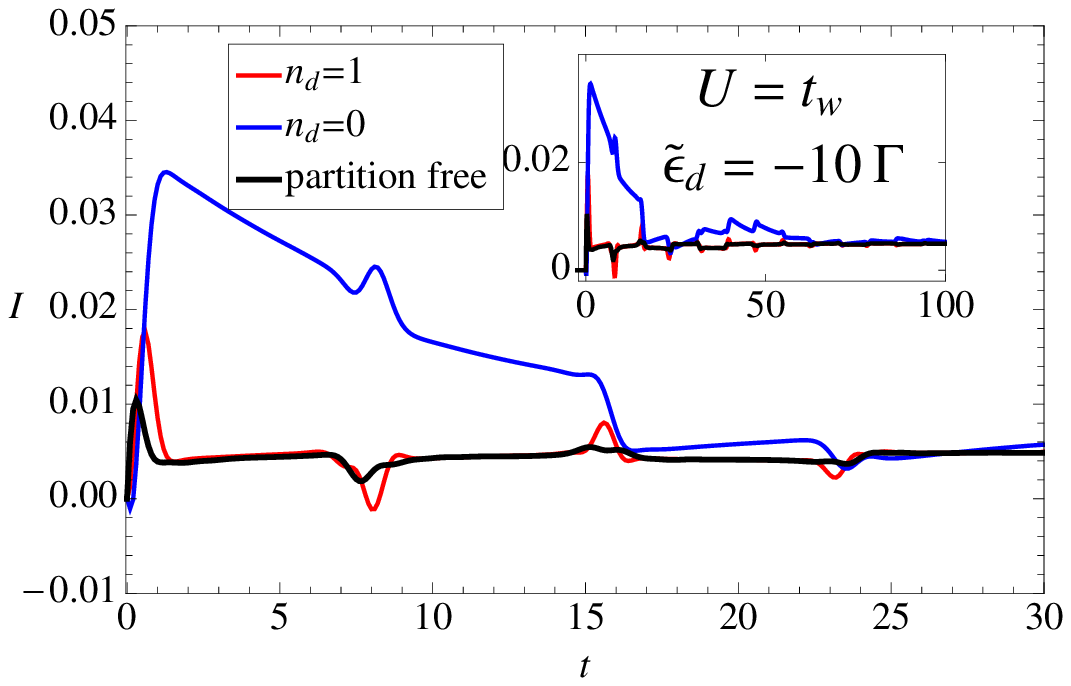}
\caption{Comparison between TD currents within the partion-free 
scheme (black) vs partitioned scheme with $n_{d}=0$ (blue) and 
$n_{d}=1$ (red). $U=0$ in the upper panel and $U-t_{w}$ in the lower 
panel.  The rest of parameters and 
units are the same as in Fig. \ref{fig1}. The insets display the TD 
result for a longer propagation time. The TD curve of the 
partition-free case has been shifted of $t_{th}=100$. }
\label{fig4}
\end{figure}

We now focus on the effects of the different initial conditions. To 
this end we compare the results obtained within the partitioned scheme vs the 
ones obtained within partition-free scheme.
In Fig. \ref{fig4} we plot the TD current
in the two schemes without (upper panel) and with (lower panel)
screening interaction $U$. For the partitioned scheme we 
employ two different initial values of the QD density, namely 
$n_{d}=1$ (red curve) and $n_{d}=0$ (blue curve).
We recall that in order to display the partition-free curves together with the partitioned 
ones, we shift the former of $t_{th}$.   
We observe that the three currents correctly reach the same steady-state 
and do it via similar long-lasting sequences of current spikes,
(typical of the FCB regime), and, as expected, the current in presence of $U$ is 
enhanced with respect to that calculated at $U=0$ (Coulomb deblocking). 
In this case the partitioning effects reflect in a quantitative 
change of the early transient current (see the black curves vs the 
red and blue ones), even 
though the qualitative FCB-like transient behavior does not change.

\begin{figure}[tbp]
\includegraphics[width=7.25cm]{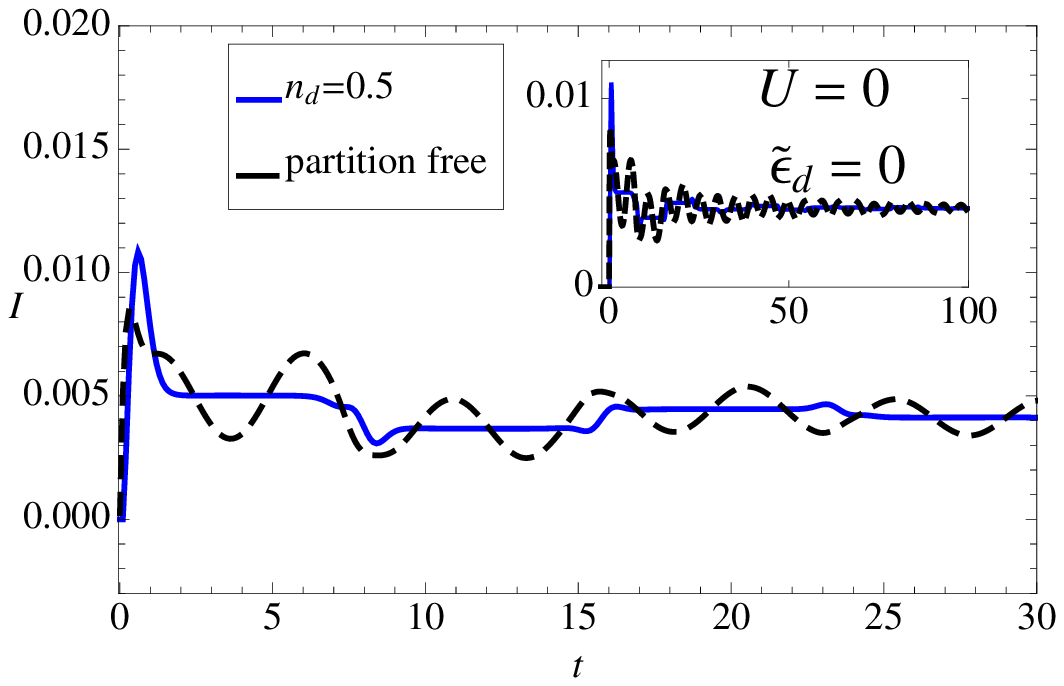}
\includegraphics[width=7.25cm]{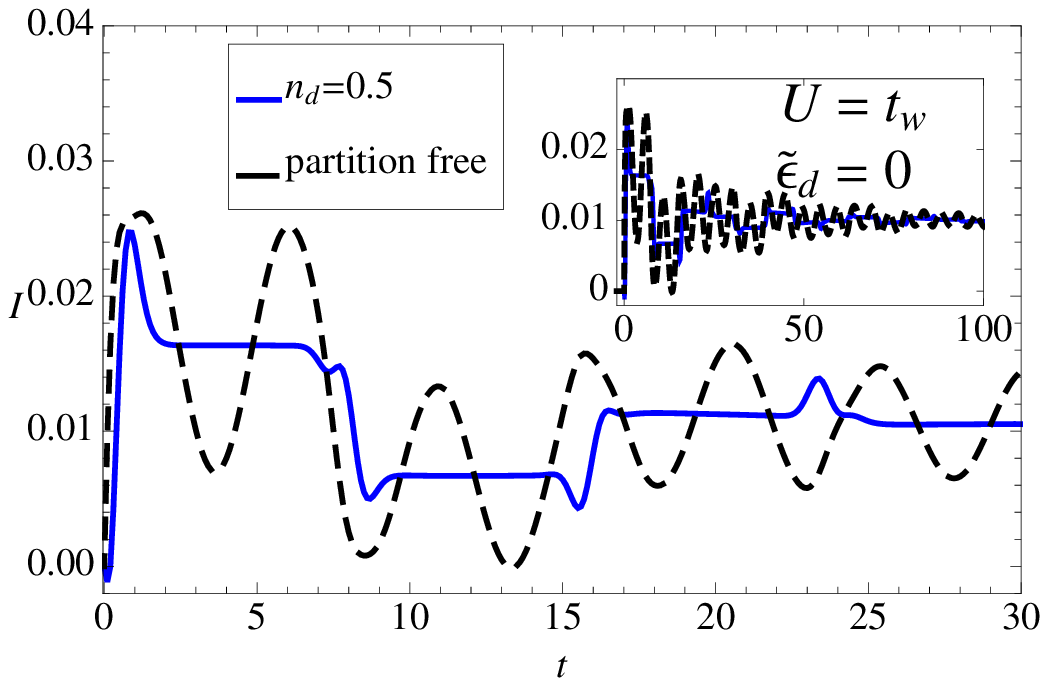}
\caption{Comparison between TD currents in the resonant
condition $\tilde{\varepsilon}_{d}=0$ within the partion-free 
scheme (dashed black) vs partitioned scheme with $n_{d}=0.5$ (blue).
$U=0$ in the upper panel and $U=t_{w}$ in the lower 
panel.  The rest of parameters and 
units are the same as in Fig. \ref{fig1}. The insets display the TD 
result for a longer propagation time. The TD curve of the 
partition-free case has been shifted of $t_{th}=100$.}
\label{fig5}
\end{figure}

\begin{figure}[tbp]
\includegraphics[width=7.25cm]{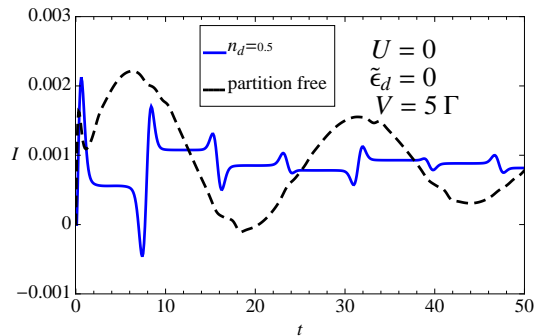}
\caption{Comparison between TD currents in the resonant
condition $\tilde{\varepsilon}_{d}=0$ within the partion-free 
scheme (dashed black) vs partitioned scheme with $n_{d}=0.5$ (blue).
$U=0$ and bias $V=5\G$.  The rest of parameters and 
units are the same as in Fig. \ref{fig1}. The TD curve of the 
partition-free case has been shifted of $t_{th}=100$. }
\label{fig6}
\end{figure}

In order to magnify the partitioning effects, we 
focus on the perfect resonant case, obtained by setting  $\tilde{\ve}_{d}=0$.
Here (see Fig. \ref{fig5}) the partition-free transient current displays
{\it qualitative differences}. If we use $n_{d}=1/2$ 
for partitioned scheme, the TD density is pinned in both schemes at the constant 
value $n(t)=1/2$ at each time (not shown), due to symmetry constraints.
Despite the TD density is identical in the two schemes, the TD 
currents are remarkably different. In particular the FCB pattern
tends to disappear in the partition-free case, and the transient 
current has a smooth oscillating behavior with dominant frequency $V/2$, i.e 
the energy difference between the molecular level and the Fermi energy 
of the leads. Here the interaction $U$ (lower panel of Fig. \ref{fig5}) amplifies these oscillations, 
thus enhancing further the difference between the two currents.
This behavior is confirmed by decreasing the bias from $V=26\G$ to 
$V=5\G$, as shown in Fig. \ref{fig6}. It is clear that the transient frequency
of the partition-free current reduces according to the smaller value 
of $V$, whereas the spikes of the partitioned current continue to 
appear with periodicity given by the phonon frequency $\w_{0}$.

This qualitative change can be interpreted as follows:
At the resonance ($\tilde{\ve}_{d}=0$) and if the leads are initially 
relaxed around the QD, the {\it dressed} tunneling (that involves the excitation 
of a phonons with energy $\w_{0}$) has the same probability as the  {\it bare} tunneling 
(that involves virtual transition between the QD and the Fermi level of 
the leads). In the partitioned case, instead, the  {\it bare} 
tunneling is suppressed because the electron charge in the leads is 
optimally rearranged in the proximity of the QD.

\section{Conclusions}

We presented a systematic study of the
time-dependent transport properties of the Anderson-Holstein model in presence 
of dot-lead screening interaction. 
Thanks to the analytic expression of the 
correlated-polaron embedding self-energy we gained a clear
understanding of how the two interactions combine together.
The validity of the approach was also corroborated by comparing our results against
exact data available in the literature.
Our approximated scheme allows for a calculation of the time-dependent density
that improves the current state-of-the-art, and at the same incorporates the screening 
effects in a physically correct way.
The transient behavior of the current was carefully analyzed by considering different
initial contacting of the leads.
We showed that at early times the current can
exhibit qualitative different behaviors, depending if the 
electron liquid in the leads is relaxed or not
around the QD before the switching of the bias.
At resonance, we found that the partition-free current displays coherent 
oscillations with frequency equal to the applied bias,
whereas the partitioned current  has
a periodicity dictated by the phonon frequency.

We acknowledge funding by MIUR FIRB 
grant No. RBFR12SW0J.


\end{document}